\documentclass[12pt]{article}
 
\begin{document}

\newcommand{\e}{\epsilon}
\renewcommand{\a}{\alpha}
\renewcommand{\b}{\beta}
\renewcommand{\k}{\kappa}
\newcommand{\p}{\partial}
\renewcommand{\phi}{\varphi}

\newtheorem{theorem}{Theorem}

\title{Passive Tracer Dispersion \\with Random or Periodic Source
\footnote{This work was supported by the National Science
                Foundation Grant DMS-9704345.} }

\author{Jinqiao Duan\\
Clemson University \\
Department of Mathematical Sciences \\
Clemson, South Carolina 29634, USA. \\
E-mail: duan@math.clemson.edu $\;\;$ Fax: (864) 656-5230}

\date{April 29, 1998  }

\maketitle

\begin{abstract}

In this paper, the author investigates the impact of external sources
on the pattern formation and long-time behavior of concentration profiles
of passive tracers in a two-dimensional shear flow.
It is shown that a time-periodic concentration profile 
exists for time-periodic external source, while for random
source, the distribution functions of
all concentration profiles weakly converge to a unique 
invariant measure (like a stationary state in 
deterministic systems) as time goes to infinity.

\bigskip


\end{abstract}

\newpage
\section{Introduction}

The dispersion of passive tracers (or passive scalars) occur
in various geophysical and environmental systems, such as
discharge of pollutants  into coastal
seas or rivers,  and temperature or salinity evolution
in oceans. Tracers are called passive when they do 
not dynamically affect the background fluid velocity field.
For the benefit  of  
better environment, it is important to understand the
dynamics of such  passive tracers.

The Eulerian approach for studying passive tracer dispersion
attempts to understand the evolution of
tracer concentration profile as a continuous field quantity 
(\cite{Clark}, \cite{Schnoor}).  

We consider two-dimensional passive tracer dispersion
in a shear flow $(u(y), 0)$ (assume that $u(y)$ is bounded).
The passive tracer concentration profile $C(x,y,t)$ then satisfies the
advection-diffusion equation (\cite{Clark})
\begin{eqnarray}
 C_t + u(y)C_x = \k (C_{xx}+C_{yy}) + \mbox{source},
  \label{eqn}
\end{eqnarray} 
where the source (or sink) term accounts for effects
of chemical reactions (\cite{Clark}), external injections
of pollutants (\cite{Schnoor}, \cite{Hunt}, \cite{Serrano}),   
or heating and cooling (\cite{Chelton}). Here $\k >0$ is the diffusivity
constant.

There has been a lot of research on the advection-diffusion
equation {\em without source}; see,
for example, \cite{Clark}, \cite{Smith}, \cite{Young},
\cite{Mezic}, \cite{Fannjiang}, \cite{Klyatskin},
\cite{Mercer} and \cite{Rosencrans}.

In this paper, we study the impact of the external sources
on the  pattern formation and long-time behavior of the concentration profile.
We assume that the  concentration profile satisfies
double-periodic boundary conditions
\begin{eqnarray}
 C, C_x, C_y  \;\; \mbox{are double-periodic in $x$ and $y$ with period $1$},
 \label{BC}
\end{eqnarray}
and appropriate initial condition
\begin{eqnarray}
 C(x,y,0) = C_0(x,y).
 \label{IC}
\end{eqnarray}
We will consider two classes of sources:
time-periodic source $f(x,y,t)$ in Section 2
and random source in Section 3. We will summarize results in
Section 4.

\section{Time-periodic source}

In this section we consider the advection-diffusion
equation with time-periodic source (\ref{eqn})
\begin{eqnarray}
 C_t + u(y)C_x = \k (C_{xx}+C_{yy}) + f(x,y,t),
  \label{eqn1}
\end{eqnarray} 
where $f(x,y,t)$ is periodic in $t$ with period $T>0$.

Integrating   both sides of (\ref{eqn1})
with respect to $x, y$ on the domain $D = [0, 1] \times [0, 1]$,
we get

\begin{eqnarray}
&  &  \frac{d}{dt} \int\int  C dxdy   + \int\int u(y)C_x dxdy  \nonumber \\
& = & \k \int\int (C_{xx}+C_{yy}) dxdy + \int\int f(x,y,t)dxdy.
\end{eqnarray} 
Note that
$$       \int\int u(y)C_x dxdy = 0      $$
and
$$      \int\int (C_{xx}+C_{yy}) dxdy = 0      $$
due to the double-periodic boundary  conditions (\ref{BC}).
We thus have 
\begin{eqnarray}
 \frac{d}{dt} \int\int  C dxdy   = \int\int f(x,y,t)dxdy.
\end{eqnarray}
Here and hereafter, all integrals are with respect to $x, y$
over $D$. 
Thus, when there is no source, the spatial average or mean of the
concentration $C(x,y,t)$ does not change with time.
When there is a source, the time-evolution of
the spatial average of    $C(x,y,t)$
is determined only by the source term.
In order to understand more delicate
impact of source on the evolution  of
$C(x,y,t)$, it is appropriate to assume that
the source has zero spatial average or mean:
\begin{eqnarray}
\int\int f(x,y,t)dxdy = 0.
\end{eqnarray}
With such a source, the mean of $C(x,y,t)$
is a constant. Without loss of generality or
after removing the non-zero constant by a translation,
we may assume that $C(x,y,t)$ has zero-mean.
So we study the dynamical behavior of
$C(x,y,t)$ in zero-mean spaces.

We use $\dot L^2_{per} (D)$ to denote the standard
function space of square-integrable double-periodic (of period $1$)
functions with zero mean. The usual norm in this
space is denoted as $\| \cdot \|$.

Note that the linear operator $-\k (\p_{xx}+\p_{yy}) +u(y)\p_x$
is sectorial (\cite{Henry}, p. 19) in $\dot L^2_{per} (D)$. Thus
if $f(x,y,t)$ has continuous derivative in time $t$,  
the linear system (\ref{eqn1}), 
(\ref{BC}), (\ref{IC}) has a unique strong solution
for every $C_0(x,y)$ in $\dot L^2_{per} (D)$
(\cite{Henry}, p. 52).  
We now show that this system  is a dissipative system in the sense 
(\cite{Krasnoselskii} or \cite{Hale}) that
all solutions $C(x, y, t)$ approach a bounded set in $\dot L^2_{per} (D)$ as
time goes to infinity. A $T-$time-periodic dissipative
system in a Banach space has at least one $T-$time-periodic
solution. This result follows from a Leray-Schauder topological degree
argument and the Browder's principle (\cite{Krasnoselskii}, p.235).    

Multipling (\ref{eqn1}) by $C(x, y, t)$ and
integrating over $D$, we get

\begin{eqnarray}
&   & \frac12 \frac{d}{dt}\|C\|^2  + \int\int u(y)C_x C dxdy  \nonumber  \\ 
& = &  - \k \int\int | \nabla C |^2 dxdy + \int\int f(x,y,t)C dxdy.
 	\label{estimate1}
\end{eqnarray} 

Note that, using the double-periodic boundary  conditions (\ref{BC}),
\begin{eqnarray}
   \int\int u(y)C_x C dxdy = 0 . 
  	\label{estimate2}
\end{eqnarray}

We further assume that the square-integral of $f(x,y, t)$ with respect to 
$x, y$ is bounded in time. 
Then, by  the Young inequality, 
\begin{eqnarray}
 \int\int f(x,y,t)C dxdy  & \leq &  \frac1{2\e} \int\int |f(x,y,t)|^2 dxdy 
		        + \frac{\e}2 \int\int |C|^2 dxdy     \nonumber  \\
	& \leq & \frac{M}{2\e}  + \frac{\e}2 \int\int |C|^2 dxdy,
		\label{estimate3}   
\end{eqnarray}
where $ M > 0 $ is a constant independent of $t$ and
$\e>0$ is an arbitrary positive number.

Since $C$ has zero mean, we can use the Poincar\'e  inequality 
(\cite{Gilbarg-Trudinger}, p. 164) to obtain
\begin{eqnarray} 
  \|C\|^2   \leq    2\pi  \| \nabla C \|^2.
      \label{estimate4}
\end{eqnarray}

Putting $(\ref{estimate2}), (\ref{estimate3}), (\ref{estimate4})$ 
into $(\ref{estimate1})$, we obtain
\begin{eqnarray} 
\frac12 \frac{d}{dt}\|C\|^2    
& \leq &  (\frac{\e}2 - \frac{\k}{2 \pi}) \| C \|^2 + \frac{M}{2\e},
	\label{estimate5}
\end{eqnarray}   
or
\begin{eqnarray}
 \frac{d}{dt}\| C \|^2      
& \leq  &  ( \e - \frac{\k}{\pi}) \| C \|^2 + \frac{M}{\e}.
\end{eqnarray}
We now fix $\e>0$ so small that $\e - \frac{\k}{\pi}<0$.
By the Gronwall inequality (\cite{Temam}), we finally get
 
\begin{eqnarray}
 \|C\|^2  \leq (\|C_0\|^2  + 
 	\frac{M}{\e( \e - \frac{\k}{\pi})} )e^{( \e - \frac{\k}{\pi})t} 
       +  \frac{M}{\e(\frac{\k}{\pi}-\e)}.
\end{eqnarray} 
Hence all solutions $C(x,y,t)$ enter a bounded
set in $\dot L^2_{per}$, 
$$
\{ C:  \;\;   \|C\| \leq  \sqrt{ \frac{M}{\e(\frac{\k}{\pi}-\e)} } \},
$$
as time goes to infinity. The system (\ref{eqn1}) is  therefore a
dissipative system and hence has at least one $T-$time-periodic
solution (\cite{Krasnoselskii}, p.235).

We thus have the following conclusion.
 
\begin{theorem}
Assume that the source $f(x, y, t)$ is time-periodic with period $T>0$
and is continuously differentiable with time $t$.  Also assume that
its mean (spatial average) is zero and its  spatial square-integral 
is bounded in time. 
Then the advection-diffusion problem with time-periodic source
\begin{eqnarray}
 C_t + u(y)C_x & = & \k (C_{xx}+C_{yy}) + f(x,y,t),  \\
 C, C_x, C_y & \;\; & \mbox{are double-periodic in $x$ and $y$ with period $1$}, \\
 C(x,y,0)     & = &  C_0(x,y),  
\end{eqnarray} 
has a time-periodic solution with period $T>0$.
\end{theorem}

We remark that it is generally difficult to show
existence of time-periodic solutions in a spatially extended
system. Our result provides such a proof of existence 
for an advection-diffusion system with source.

\section{Random source}

In this section we consider the passive tracer dispersion problem
(\ref{eqn}) with a white noise source. We want to study
the long-time behavior of the distribution function of the
random concentration profile $C(x,y, t)$.
A white noise    is usually modeled
by the Ito derivative of a space-time $Q-$Wiener process $W(x,y,t)$
which has zero mean value (expectation) for each $t$. 
$Q$ is a symmetric non-negative linear operator
in $\dot L^2_{per}(D)$; see \cite{DaPrato}, \S 4.1.
In this case, (\ref{eqn}) becomes

\begin{equation}
 dC  = (- u(y)C_x + \k (C_{xx}+C_{yy}))dt + dW.   
   \label{eqn2}
\end{equation} 
This is a linear stochastic differential equation
in $\dot L^2_{per}(D)$.  As we mentioned in the last section,
$\k (\p_{xx}+\p_{yy}) - u(y)\p_x$ generates an analytic
semigroup $S(t)$ in $\dot L^2_{per}(D)$.

Assume that $Q$ satisfies 
 
\begin{equation}
 \int_0^t Tr\; S(r) Q S^{\star} (r) dr < +\infty.
 	\label{condition}
\end{equation}
Then, as can be shown in \cite{DaPrato}, \S 5.2 and \S 5.4,
for every initial condition $C_0(x, y)$ in $\dot L^2_{per}$, 
there exists a unique global mild solution $C(x,y, t)$ of the stochastic  
differential equation (\ref{eqn2}) under (\ref{BC})  and (\ref{IC}).

The corresponding deterministic  equation for (\ref{eqn2}) is
\begin{eqnarray}
  C_t + u(y)C_x = \k (C_{xx}+C_{yy}).
\end{eqnarray}
As in (\ref{estimate5}), we have
\begin{eqnarray} 
 \frac{d}{dt}\|C\|^2  \leq  - \frac{\k}{ \pi}  \| C \|^2  ,
	\label{estimate6}
\end{eqnarray}
Thus, by the Gronwall inequality,
\begin{eqnarray}
 \|C\|^2  \leq  \|C_0\|^2 e^{- \frac{\k}{\pi} t}. 
\end{eqnarray} 
Therefore,
$$
lim_{t\rightarrow +\infty} S(t)C_0(x,y) 
	= lim_{t\rightarrow +\infty} C(x,y,t) =0,
$$
which implies that, 
$$
lim_{t\rightarrow +\infty} \|S(t)\| = 0.
$$
Therefore, according to Theorem 11.11 in \cite{DaPrato},
the stochastic differential equation
(\ref{eqn2}) has a unique invariant measure
(like a stationary state for a deterministic
partial differential equation), and 
the distribution function of
any other solution process $C(x,y,t)$
weakly converges to this invariant measure in $\dot L^2_{per}$
as $ t\rightarrow +\infty $.

We thus have the following result.

\begin{theorem}
Assume that $Q-$Wiener process $W(x,y,t)$ satisfies 

\begin{equation}
    \int_0^t Tr\; S(r) Q S^{\star} (r) dr < +\infty.
\end{equation}   
Then the advection-diffusion problem with random source

\begin{equation}
      dC  = (- u(y)C_x + \k (C_{xx}+C_{yy}))dt + dW   
\end{equation}
has a unique invariant measure in $\dot L^2_{per}$, 
and the distribution functions of
all other solutions weakly converge to  this unique invariant measure
as $ t\rightarrow +\infty $.
\end{theorem}

\section{Discussions}

In this paper, we have studied the impact of external sources
on the  pattern formation and long-time behavior of concentration profiles
of passive tracers in a two-dimensional shear flow.
We have shown that a time-periodic concentration profile 
exists for time-periodic external source, while for random
source, the distribution functions of
all   concentration profiles weakly converge to a unique 
invariant measure (like a stationary state in 
deterministic systems) as time goes to infinity.

\bigskip

{\bf Acknowledgement.} The author thanks Jim Brannan for
			useful comments.

\end{document}